\documentclass[prl,twocolumn,showpacs,superscriptaddress]{revtex4}
\usepackage{graphicx}
\usepackage{dcolumn}% Align table columns on decimal point
\usepackage{bm}% bold math

\begin{document}

\title
{Numerical study of vortex system quantum melting}

\author{M. V. Zyubin }
\author {I. A. Rudnev}
\email{rudnev@supercon.mephi.ru}
\author {V. A. Kashurnikov }
\affiliation{Moscow Engineering Physics Institute, Moscow, 115409
Russia}

\date{\today}
\begin{abstract}
We report a numerical study of the vortex system in the two
dimensional II-type superconductors. We have proposed a
phenomenological model that takes into account quantum fluctuations
of Abrikosov's vortices. The results of the quantum Monte-Carlo
simulations by the SSE algorithm show that the thermal fluctuations
are dominated by quantum fluctuations at low temperatures. In
particular, we demonstrate the possibility of the quantum melting
transition for vortex system in the temperature region where
thermal melting transition is improbable.
\end{abstract}

\pacs{74.70.Kn, 74.25.Qt, 05.10.Ln}

\maketitle

\section{Introduction}
Physics of the vortex systems is a very attractive field of the
modern condensed matter physics due to various phases and phase
transitions associated with vortex matter. Nature of the phase
transitions in the vortex system is a subject of numerous both
theoretical and experimental studies (see, for example, the reviews
\cite{Blatter, Brandt}). It is well known that at low temperatures
vortex system is in solid phase. Structure (quenched) disorder or
other kinds of fluctuations have an influence on vortex system
state. Thus as the thermal fluctuations increase the vortex lattice
melts and system undergoes transition from solid to the liquid
phase. Some peculiarities of vortex lattice melting at the presence
of structure defects have been obtained by numerical simulations
and were discussed in Refs. \cite{M1,M2,M3,M4,R1,R2,R3,R4,R5}.

Resent experimental results demonstrate an alternative mechanism of
the order destruction of vortex systems. This mechanism is
associated by the some researches with quantum fluctuations in the
vortex system \cite{Q1,Q2,Q3,Q4,Q5}. So the authors of the paper
\cite{Q5} investigated magnetic relaxation in $MgB_{2}$ thin film
and made a conclusion on the presence of the quantum creep of the
vortices at $T \rightarrow 0$. They supposed that the creep is
induced by quantum fluctuations. The transport properties
investigations of $k-(BEDT-TTF)_{2}Cu(NCS)_{2}$ organic
superconductor \cite{Q2} as well as the measurements of dc and ac
complex resistivities for thick $\alpha-Mo_{x}Si_{1-x}$ films
\cite{Q3} indicated, in the author's opinion, a presence of quantum
fluctuations too. Note that the quantum fluctuations are supposed
to be essential for thin superconducting films and high-layered
superconductors, i.e. for systems with reduced dimensions. The
quantum fluctuations became potentially significant at low
temperatures $T\rightarrow 0$, where the thermal fluctuations are
negligibly small. In other words, quantum fluctuations can give
rise to melting of the vortex system in the temperature range where
the vortex system is expected to be in solid phase from the
classical point of view.

Though there are many experimental researches of the quantum
fluctuations influence on properties of the vortex system,
simulations of vortex system with quantum fluctuations are
practically absent in literature. One exception is the
paper\cite{Sim} where study of two dimensional vortex system has
been done. However world-line Monte-Carlo algorithm in discrete
imaginary time used in \cite{Sim} did not allow the authors of the
paper to simulate the processes of creation-annihilation of
vortices in external magnetic field.

In this Letter we propose a phenomenological model for description
of quantum fluctuations in the vortex system and report simulation
results which were obtained by the loop quantum Monte Carlo
algorithm in continuous imaginary time.

\section{ Model}

To describe the quantum fluctuations of vortex system we propose to
consider the vortices as hard-core bosons and introduce quantum
tunneling term into Hamiltonian. So, Hamiltonian of vortex system
takes the form:
\begin{eqnarray}
\hat{H} = \label{Ham}
-t\sum_{<i,j>}a^{+}_{i}a_{j}+h.c.+\frac{1}{2}\sum_{i\neq
j}V(r_{i,j})n_{i}n_{j} \nonumber\\-\sum_{i}{h_{i}n_{i}},
\end{eqnarray}

where the first term describes the quantum tunneling of the
vortices between sites $i j$ with hoping amplitude $t$ - the
phenomenological parameter for describing of the wave function
overlap. In the general case vortex can hope between any different
sites, but for the simplicity of the model we take into account the
hopping between only nearest neighboring sites. The second term
describes a pairwise interaction between vortices. And the third
term includes all linear interactions, namely vortex self-energy,
interaction between vortices and pinning centers, interaction
between vortices and external magnetic field. Quantum behavior of
the vortex system is described by the first term. The second and
third terms are the same as for classical vortex system (see, for
example Ref. \cite{R3}). In this paper we choose:

\begin{figure}
\begin{center}
\resizebox{85mm}{!}{\includegraphics{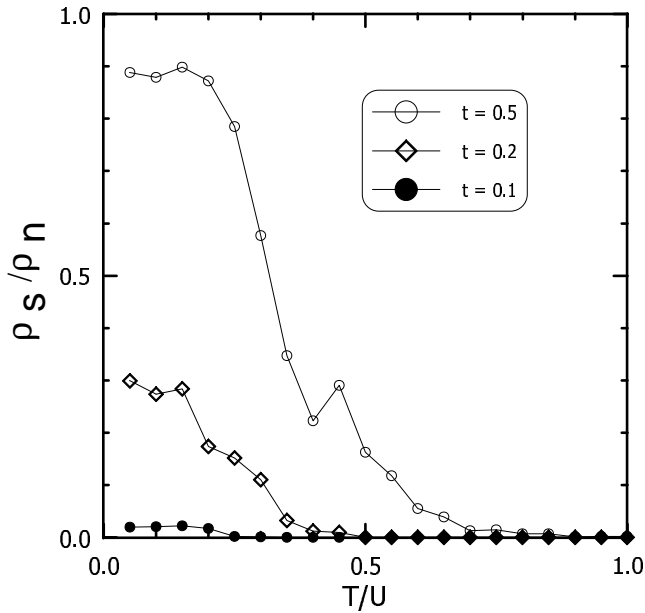}}
\end{center}
\caption{\small Superfluid density versus temperature for different
hopping amplitudes $t$ at $U=1.0$, $h=0.5$.} \label{f1}
\end{figure}

\begin{equation}
V(r_{i,j}) = UK_{0}(r_{i,j}/\lambda),
\end{equation}

\begin{equation}
h_{i} = h+\mu_{i},
\end{equation}
where $h$ - external field, $\mu_{i}$ - pinning potential, $K_{0}$
- Bessel function, $\lambda$ - penetration depth, $r_{i,j}$ - a
distance between the vortices at sites $i$ and $j$.

A study of the system described by equation (\ref{Ham}) is a very
complicated problem. Analytic solutions of the model \ref{Ham}) are
do not available. Therefore it is necessary to use simulation
methods. The quantum Monte Carlo algorithms are the most powerful
tools for exploring quantum many body systems, such as a system
(\ref{Ham}). Quantum Monte Carlo methods allow us to calculate
thermodynamic characteristics of the model and obtain visual
qualitative picture of complex physical processes
\cite{Hirch,Scal,Evertz,Sand3,MySSE}.

\begin{figure}
\begin{center}
\resizebox{85mm}{!}{\includegraphics{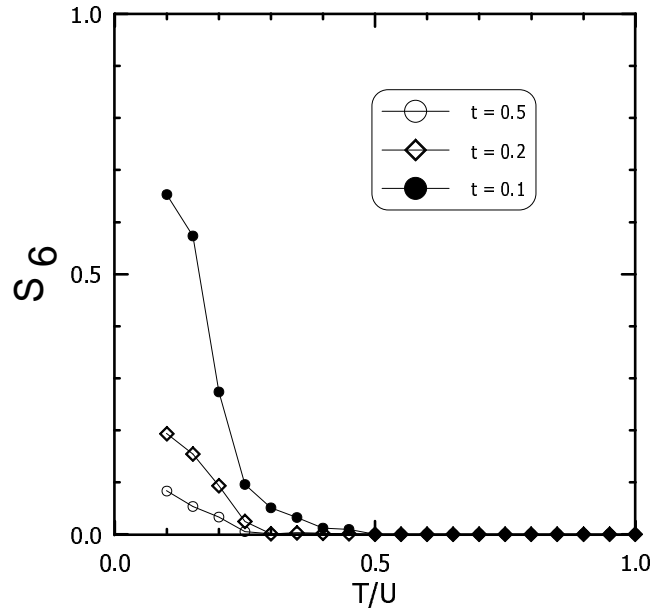}}
\end{center}
\caption{\small Structure factor $S_{6}$ versus temperature for
different hopping amplitudes $t$ at $U=1.0$, $h=0.5$. } \label{f2}
\end{figure}

Note that by use of equation (\ref{Ham}) we can describe both
quantum and classical regimes of the vortex system.

To analyze the model (\ref{Ham}) we use SSE (stochastic series
expansion) algorithm modified for long-range potential. The
algorithm is an exact numerical method and allows us to simulate
quantum many body system in wide region of external fields and
temperatures \cite{MySSE}. The method can be implemented in the
both site \cite{Sand3,MySSE} and interaction representations
\cite{SSEInt}. In the case of long-range potential an interaction
representation is preferable. For the interaction representation
Hamiltonian splits into a diagonal part $\hat{D}$ and a
perturbation part $\hat{V}$\cite{Book,SSEInt}:

\begin{equation}
\hat{H} = \hat{D}+\hat{V}.
\end{equation}
The partition function takes the form:
\begin{eqnarray}
Z =
\sum^{\infty}_{n=0}(-1)^{n}\int^{\beta}_{0}d\tau_{1}\int^{\tau_{1}}_{0}d\tau_{2}...
\int^{\tau_{n-1}}_{0}d\tau_{n}\nonumber\\
Tr\{\hat{V}(\tau_{1})\hat{V}(\tau_{2})...\hat{V}(\tau_{n})\},
\end{eqnarray}
where
\begin{equation}
\hat{V}(\tau) = e^{\tau \hat{D}}\hat{V}e^{-\tau \hat{D}}.
\end{equation}

After that we choose basis $\{|\alpha\rangle\}$ where $\hat{D}$ is
diagonal and decompose the perturbation part into bond operators:
\begin{equation}
\hat{V} = \sum_{b}\hat{H_{b}},
\end{equation}
where $b$ - bond index. So we can rewrite the partition function in
the following way \cite{SSEInt}:
\begin{eqnarray}
Z =
\sum_{\alpha}\sum^{\infty}_{n=0}\sum_{T_{n}}\int^{\beta}_{0}d\tau_{1}\int^{\tau_{1}}_{0}d\tau_{2}...
\int^{\tau_{n-1}}_{0}d\tau_{n}W(\alpha,T_{n},\tau),\nonumber\\ W =
(-1)^{n}(e^{-\beta E_{0}}\prod^{n}_{p=1}e^{-\tau_{p}
(E_{p}-E_{p-1})})\langle\alpha|\prod^{n}_{p=1}\hat{H}_{b_{p}}|\alpha\rangle,
\end{eqnarray}

where $E_{p} = \langle\alpha(p)|\hat{D}|\alpha(p)\rangle$, $T_{n}$
a sequence of non-diagonal operators. To simplify simulations we
add unit operators into the operator sequence $T_{n}$ and obtain
the operator sequence $T_{m}$. The Monte Carlo simulation is
carried out by diagonal and loop updates. The simulation starts
with an arbitrary state $|\alpha\rangle$ and the operator sequence
$T_{m}$ containing only unit operators. The diagonal update
contains one attempt to interchange diagonal and unit operators.
Note that the diagonal update changes the expansion power $n$ by
$\pm1$. In a stage of the loop update the interchange of diagonal
and non-diagonal operators is carried out with the fixed expansion
power $n$. So the loop update provides the creation-annihilation of
kinks. At the same time the state of the system $|\alpha\rangle$
can be changed. The method allow us to operate directly with
continuous imaginary time $\tau$ and grand canonical ensemble.

\section{Results}
We consider a two-dimensional vortex system on triangular grid with
periodic boundary conditions. To analyze behavior of the system we
calculate a structure factor
\begin{equation}
S_{6} = \langle\sum^{N}_{i=1} \frac{1}{Z_{i}} \sum^{Z_{j}}_{j=1}
e^{i6\theta_{ij}}\rangle,
\end{equation}
and so-called superfluid density
\begin{equation}
\label{ps} \rho_{s} = \frac{1}{2t\beta}\langle W^{2}\rangle.
\end{equation}

Superfluid density is introduced in according to the conventional
determination \cite{Batrouni}. Superfluid density depends on the
mean square of winding number that characterizes a topological
configuration of the world lines. Note that the nonzero value of
$\rho_{s}$ means the presence of the energy dissipation as a result
of the vortex quantum tunneling, i.e. there is a rise of
resistivity. And on the contrary as $\rho_{s}\longrightarrow0$
quantum creep tends to zero and the only thermal dynamics of
vortices is present (in case of current absent).

Calculations have been done for the vortex system on 10 x 10
triangular grid with the periodic boundary conditions. The grid
spacing was chosen to be $0.25\lambda$. One pinning center with the
potential $\mu = 0.1 U$ was introduced into the system to fix a
vortex lattice. Two sets of calculations have been done. The
temperature is changed at the fixed hopping amplitude $t$ and vice
verse hopping amplitude is changed $t$ at the fixed temperature.

Fig. ~\ref{f1} shows the superfluid density versus temperature at
different hopping amplitudes $t$. As seen from dependencies at
small hopping amplitude $t=0.1$ superfluid density is almost absent
in whole range of temperatures. While at $t = 0.5$ well-defined
plateau arises on graph. The plateau corresponds to the quantum
regime of vortex system. At the same time superfluid density drops
rapidly as the temperature increase. This corresponds to
transitions of the system to classical regime. As mentioned above,
the superfluid density has another meanning for vortex system then
usual one for hard-core boson system. Generally if the physical
system is in superfluid state the dissipation is absent. However,
as well known, high vortex mobility is accompanied by energy
dissipation and therefore high value of superfluid density
corresponds to high energy dissipation. So quantum fluctuations can
give rise to dissipation in the temperature range where there is no
dissipation in the classical approach.

The structure characteristics of vortex system along with the
superfluid density are of interest. Figure 2 shows the dependencies
of the structure factor $S_{6}$ on temperature at different hopping
amplitudes $t$. As seen from the curves the structure factor tend
to zero at high temperatures for all considered values of hopping
amplitude $t$. As the temperature decreases the structure factor
increases. Moreover, in the almost classical case $t = 0.1$ the
system is more ordered at low temperatures then in the case of
strong quantum fluctuations $t = 0.5$.

\begin{figure}
\includegraphics[width=8.5cm]{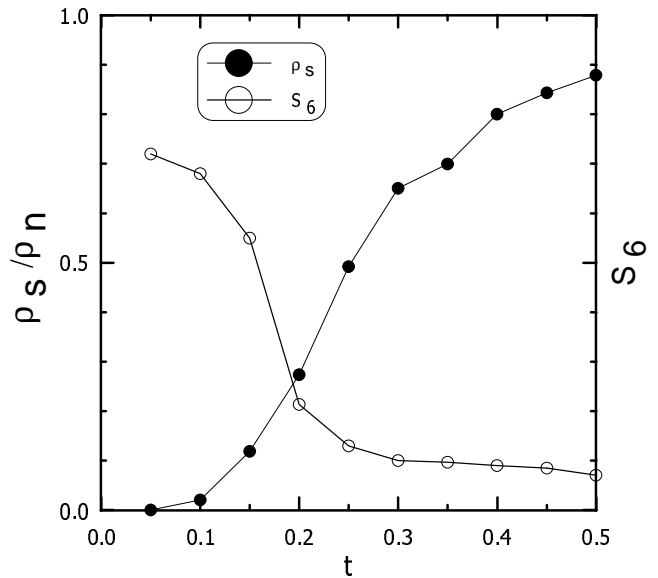}
\caption{\small Superfluid density and structure factor $S_{6}$
versus hopping amplitude $t$ at temperature $T = 0.1U$, $U=1.0$,
$h=0.5$.} \label{f3}
\end{figure}

To examine a dynamics of the system with the increase of the
amplitude of quantum fluctuation, we have done calculations at
fixed temperature $T=0.1 U$ and different values of hopping
amplitude $t$. As seen from Fig. ~\ref{f3} an enhancement of the
hopping amplitude $t$ leads to the increase of the superfluid
density. The last one corresponds to the transitions of the system
into the quantum regime. Simultaneously the structure factor
$S_{6}$ decreases, i.e. the vortex system disordering takes place.
In other words the vortex system melts under quantum fluctuations.

\section{Summary}

We have numerically examined the behavior of the Abrikosov's
vortices system in a two dimensional superconductor taking into
consideration quantum fluctuations. Simulations have been done in
wide range of temperatures at different intensity of quantum
fluctuations. We have demonstrated that the enhancement of quantum
fluctuations results in the quantum melting of the vortex system.
Note that the quantum fluctuations must be essential in thin
superconducting films, organic superconductors, high-layered
superconductors and can be observed in magnetic and transport
experiments at low temperatures.

We acknowledge financial support from RFBR under Grant No.
03-02-16979 and Russian Program "Integration" under Grant No.
B-0048.

\end{document}